\documentclass[twocolumn]{aastex62}
\usepackage{natbib}
\usepackage{graphicx}
\usepackage{multirow}
\usepackage{txfonts}
\begin{document}
\shorttitle{Outflow Induced Soft Lags}
\shortauthors{Patra, Chatterjee, Dutta, Chakrabarti \& Nandi}
\title{Evidence of Outflow Induced Soft Lags of Galactic Black Holes}
\correspondingauthor{Arka Chatterjee}
\email{arkachatterjee@bose.res.in, arka019icsp@gmail.com}
\author{Dusmanta Patra}
\affiliation{Indian Centre for Space Physics, 
 43 Chalantika, Garia St. Rd., Kolkata, 700084, India.}
\author{Arka Chatterjee}
\affiliation{S. N. Bose National Centre for Basic Sciences, 
 Salt Lake, Kolkata, 700106, India.}
\author{Broja G. Dutta}
\affiliation{Rishi Bankim Chandra College, 
 Naihati, West Bengal, 743165, India.}
\affiliation{Indian Centre for Space Physics, 
 43 Chalantika, Garia St. Rd., Kolkata, 700084, India.}
\author{Sandip K. Chakrabarti}
\affiliation{S. N. Bose National Centre for Basic Sciences, 
 Salt Lake, Kolkata, 700106, India.}
\affiliation{Indian Centre for Space Physics, 
 43 Chalantika, Garia St. Rd., Kolkata, 700084, India.}
\author{Prantik Nandi}
\affiliation{S. N. Bose National Centre for Basic Sciences, 
 Salt Lake, Kolkata, 700106, India.}

\begin{abstract}
The nature of lag variation of Galactic black holes remains enigmatic mostly 
because of nonlinear and non-local physical mechanisms which contribute to 
the lag of the photons coming from the region close to the central black holes. 
One of the widely accepted major sources of the hard lag is the inverse Comptonization 
mechanism. However, exact reason or reasons for soft lags is yet to be identified. 
In this paper, we report a possible correlation between radio intensities of several 
outbursting Galactic black hole candidates and amounts of soft lag. The correlation 
suggests that the presence of major outflows or jets also change the disk 
morphology along the line of sight of the observer which produces soft lags.

\end{abstract}


\section{Introduction}
X-ray lags of Galactic black hole candidates (hereafter GBHs) are reported in the literature 
\cite{Mi88} for quite some time. Lags are obtained using cross Fourier spectrum of 
X-ray light curves having various energy bands. The reference or soft band are selected
based on the lower cutoff of the detector and on the specificity of the object needed to be studied.
For RXTE, the soft band 2-5 keV is a good choice to study the GBHs as the Keplerian disk flux 
(\cite{ss73}) maximizes within $0.1-10$ keV. Harder photons, energized via inverse 
Compton process (\cite{ST80}), in the energy bands 5-10, 10-18 keV and more, 
are used to create cross correlate with the soft band to generate phase lag spectra. 
Depending on the sign of the lag integrated over the a particular frequency range, 
one can infers the arrival time delay between either of the two bands. Hard lags or positive 
lags occur when hard photons arrive later than their softer counterpart. The reverse is true
for soft lags (please see \cite{Fa09} and \cite{Ut14} for details on lags related 
to black holes). It is believed that inverse Comptonization always produces hard lags (\cite{P80}). 
However, GBHs exhibit complex lag behavior which are very difficult to interpret within 
the standard framework. It was understood that the lags are originated due to the 
accretion flows. 

The accretion disks around GBHs are formed due to the matter supplied from its binary 
companion. The companion star can supply matter via Roche Lobe overflow, tidal disruption 
event or winds. During the accretion, the matter gains kinetic energy by falling into 
deeper gravitational potential well of a black hole. The process heats up the matter 
which finally leads to the radiations in the entire EM spectra. The observed spectra of 
black hole binaries in X-ray regime contains two major components. Thermal black body part 
which is believed to be originated from Keplerian disk (\cite{ss73}) and a 
power law part (\cite{ST80}) generated due to inverse Comptonization of the soft photons 
inside a hot corona or Compton cloud. The origin and configuration of this 
so called Compton cloud remained mischievous until the mid to late 90’s. Several models 
like TCAF \citep{CT95}, ADAF \citep{NY94}, RIAFs \citep{YQN03} 
are proposed to explain the spectral properties. Previous to that, 
\cite{Ch90} elaborately presented the mass independent 
analytical solutions of accretion flow properties around black holes. The transonic solution 
clearly favored the shock transition as the flow can enter into a higher entropy branch during 
its course towards the black hole. Due to this sudden velocity change, the sub-keplerian matter
heats up more which creates the Compton cloud. The shock transition radius or shock 
location ($X_s$) depends 
on the angular momentum and energy of the matter at outer boundary. 
In order to understand the progress of Low Frequency QPOs during rising and declining phase of
any outburst, one may also consider advective flow models proposed by \cite{CT95};
\cite{es97} or \cite{BB99}. Most of the advective flow models predict a variable
Compton cloud which often produced by the shock formation near the centrifugal barrier
\cite{Ch99}. Numerical simulations performed by \cite{Mo94} indicated that the luminosity
variation due to shock oscillation could result in the formations of LFQPO. 
Another school of thought presented numerical simulations (\cite{FA05}; \cite{fr07} and 
\cite{in09}) which support the claim on possible origin of LFQPOs due 
to Lense-Thirring precision of disk plane with axis of rotation of the central 
black hole. However, formation of LFQPOs by Lense-Thirring precision in presence of 
viscosity and radiative cooling for relatively larger ($50~r_g$) Compton cloud, 
are yet to be tested.

Several (\cite{Cu99b}; \cite{No99}; \cite{Pou99}; \cite{Ka13}; 
\cite{Du16} (hereafter DC16), \cite{Ch17b} (hereafter CCG17b)) attempts were made 
to understand the X-ray lag spectra in presence of Comptonization, reflection and magnetic 
field and gravitational focusing. DC16 pointed out that the soft lags are seen in high 
inclination angle ($\theta_{obs}\ge 60^\circ$) sources. It was found that above a certain QPO
frequency ($\nu_{tr}$), the lag changes sign from hard to soft. 
Coupled with Ray-Tracing (\cite{Ch17a}, 
hereafter CCG17a) and Monte-Carlo simulations, CCG17b showed the energy 
dependent lag variations as a function of the size of the Compton cloud, accretion
rate and inclination angle. Effects of inclination angle (see CCG17a for further 
details) on lag magnitude were demonstrated as well. Inverse Comptonized photons 
naturally lag behind soft photons that are coming directly to the observer. 
With increasing inclination, number of reflected photons by disk increases 
(see CCG17a) also. The simulations by CCG17a showed a similar lag variation as 
were observed by DC16 in case of GX 339-4. \cite{van17} performed a survey over 
15 GBHs and also found that the soft lags were seen only in case of high $\theta_{obs}$ 
candidates. They argued that the soft lags computed for type-C QPOs are originated 
by the Lense-Thirring precision of the inner disk in cases of the high inclination 
angle Galactic black holes. 

In this picture, origin of type-B QPOs differs from type-C QPOs. B-type QPOs 
are considered to be originated from jets \citep{van17} as the amplitude of such QPOs decreases with increasing 
inclination angle. It is also worthwhile to note that the LFQPOs also change their types 
(A, B or C), centroid frequency ($0.01 \mathrm{Hz}\le \nu_c \le 20~\mathrm{Hz}$), amplitude and {\it rms} 
as an outburst progresses and very few are observed at highest/softest possible state where 
the effect of thermal Comptonization is found to be the least. 

\begin{table*}
\centering
\caption{\bf Details of Source Information and Observation}
\normalsize
\begin{tabular}{l c c c c }
Information                     & XTE 1550-564              & XTE 1859+226                &  H 1743-322        & GRS 1915+105            \\
\hline                                                                      
\hline                                                                      
Distance (kpc)                  & $4.4\pm0.5$               & $8\pm3$                     &  $10.4\pm2.9$      &  $8.6\pm2.0$            \\\hline
Mass ($M_{BH}/M_{\odot}$)       & $10.4\pm2.3$              & $10.8\pm4.7$                &  $11.4\pm0.9$      &  $12.4\pm2.0$           \\\hline
Inclination (degrees)           & $67\pm10.0$               &  $74\pm4.0$                 &  $69\pm3.0$        &  $70\pm2.0$             \\\hline
Outburst Year                   & 1998                      & 1999                        & 2003               & 1996                    \\\hline
MJD span                        & $51065-51092$             &  $51464-51480$              & $52730-52800$      & {\bf$50350-50600$}           \\\hline
Number of radio                 & 12 (843 MHz), 7 (8.6 GHz) & 11 (8.4 GHz)                & 28 (8.4 GHz)       & 35 (8.4 GHz)            \\
Observation                     &                           &                             &                    & 51 (15.2 GHz)           \\\hline
Radio Telescope                 & MOST$^\dag$, ATCA$^\ddagger$ & VLA                      & VLA                & GBI$^\star$, Ryle$^\star$  \\\hline
Project Code                    &                           & AH0669                      &AL0586, AR0508,AS0762&                         \\\hline
                                &                           &                             &AJ0302, AB1075,AR0523&                         \\\hline
                                &                           &                             &AM0773, AM0759       &                          \\\hline
X-Ray Observation               & 16                        & 65                          & 31                 & 72                      \\
(RXTE PCA)                      &                           &                             &                    &                         \\\hline
References                      & 1, 5, 6                   & 1, 2                        & 1, 3               & 1, 4, 7                   \\
\hline
\hline
\end{tabular}\\
1. \cite{Ta16} and references therein; 2. \cite{Co11}; 3. \cite{Co05}; 4. \cite{Re14}; \\ 
5. \cite{Wu02}$^\dag$; 6. \cite{Ha01}$^\ddagger$; 7. \cite{Mu01}$^\star$.

\end{table*}

As a natural
consequence of accretion, commonly, advective flow models produce significant 
amount of outflows. In presence of magnetic field \citep{Lo76}, to some extent, 
these outflows may also be collimated into compact or blobby jets depending 
on the spectral state (\cite{Fe04}; \cite{Fe09}) of the outbursting candidate. 
In recent years, the correlations between X-ray flux ($F^X$) and radio flux ($F^R$) for numerous black hole candidates 
(\cite{Me03}; \cite{Ha98}; \cite{Co03}, \cite{Ca07}, \cite{SF11}; \cite{Co11}; 
\cite{Jn12} and \cite{IZ18}) suggested a strong radiative and morphological connection 
between X-ray emitting accretion disk and synchrotron dominated radio jets. 

\begin{figure*}
\vskip 0.0cm
{\includegraphics[height=1.6\columnwidth,width=2.3\columnwidth]{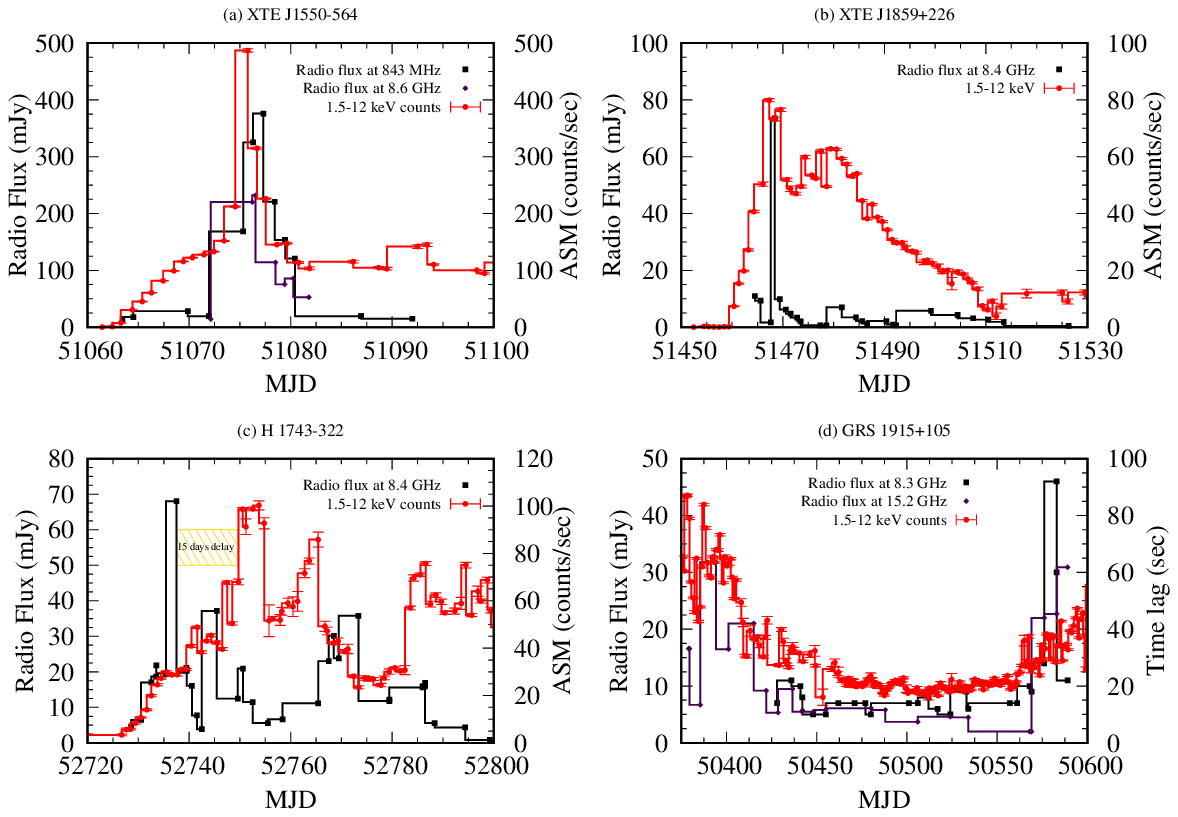}}\hskip 0.2cm
\caption{Radio flux and ASM counts (2-10 keV) are plotted for four candidates. In Fig. (a) 
and (b), we see the radio peaks are matching with the ASM maxima. In case of Fig. (c), a 
delay between radio and X-ray can be noticed (see \S 3.3 for further details). For GRS 1915+105, 
in between MJD 50350 to 50600, the radio flux has increased substantially in a coherent manner
with the X-ray counts.} 
\label{}
\end{figure*}

It is evident that the outflows/jets somehow contribute to the observed quantities like 
total flux and QPO frequency. Thus, to be able to understand the disk-jet connection and 
its implication on QPO formation, one needs to have simultaneous multi-wavelength observations. 
In the last decade, a few efforts were made to acquire a deeper understanding of physical 
mechanisms around the jet base by connecting radio, optical, and X-ray bands. \cite{Ga08} 
showed the optical lag of GX-339-4 with respect to the X-rays and invoked a possibility
of the lag to be originated by modulation of magnetic fields near the jet base. Following that,
\cite{Vi18} investigated IR to X-ray lags of GX 339-4. However, the radiative 
coupling between disk and jet which may have affected the sub-second IR to X-ray lag remained 
inconclusive from their study. Later, in 2015, V 404 Cygni had undergone an outburst and was 
well observed in wide band. \cite{Te19} claimed that the jet behavior changed from 
blobby ejecta to more compact in nature in the later stages of the outburst. The synchrotron 
break frequency of this object was found to be $\sim 2-5\times10^{14}$ Hz which is in the 
NIR-optical band. Due to the very high break frequency, \cite{Te19} also argued that
a large fraction of photons originated in the jet region might have inverse Comptonized in the hot 
corona and contribute in the observed X-ray. The positive lag (soft/negative lag in our nomenclature as 
the lesser of the two frequency bands is lagging) of $\sim 12$ minutes between 26 and 5 GHz 
indicates light crossing time of $1.5$ AU. But, the absence of any such lag in between 26 and 7 GHz bands 
complicates the scenario more.  Although their work revealed a great deal of 
information on the 2015 outburst of V404 Cygni, it did not probe the inner regions of jet base 
using X-ray timing properties. \cite{AM15} analyzed several outbursts of a popular 
black hole binary GX 339-4. The trail of hardness-phase lag followed a similar pattern to that of 
Cyg X-1 as they both belong to intermediate inclination $30^{\circ}<\theta_{obs}<50^{\circ}$ 
angle group. The soft photons (2-5 keV) for GBHs are majorly generated by the Keplerian disk
and viscous time scale of such disk is believed to be in the order of few days. 
So, variabilities found in sub-second regime are well known to be originated by the variations of 
corona or Compton cloud (for further details please see \cite{cm00}). But, it is 
intriguing to find the average lags, even in mHz domain, are dominated by the corona and jet.
\cite{Ve17} and \cite{Ma18} tried to model lag spectra between optical
and X-ray using jet models. A more recent correlation between time lags and photon-index was 
reported by \cite{Re18}. The correlation remains steady for rising state of 
low and intermediate inclination angle candidates but changes the slope for high 
inclination group (Fig. 3, \cite{RK19}) and finally inverts after photon index 
reaches 2.0. Unlike Cyg X-1, the high inclination angle GBHs have shown negative/soft lag
in X-ray band. So, the correlation might indicate about the change of lag values due to 
spectral state change, but, does not address how the morphology of disk-jet system might 
have generated the soft lags. \cite{RK19} further investigated the lag-photon
index correlation using simple jet model where optical depth of the corona/jet base falls 
off with increasing size of the jet base. The simulated data fits observational data of 
low and intermediate inclination group well. But, the jet model alone used in their 
simulation might not be adequate to simulate soft lags. So, under these situation, 
the disk-jet connection can not be ignored to understand the lag behavior of black 
hole candidates. 

\cite{Mu01} first reported that the lag sign changes in presence of higher activity in radio waves in GRS 1915+105.
However, no follow-up observation or correlation was established to conclusively add
outflows among the contributors of soft lag. More recently, CCG17b first pointed out the 
possibility of such correlation between time lag and radio flux seen in case of XTE J1550-564 
in its 1998 outburst. In the present article, we conduct a study on four GBHs where the 
correlations between radio fluxes and soft lags are investigated.

This {\it article} is structured in the following way: the next Section deals with the 
observations and data analysis. In \S 3, we examine results of each candidate along with a 
brief description of each source. Simulated energy dependent time lags in presence of 
outflows are discussed in \S 4. We discuss possible physical explanations to corroborate 
observational results in \S 5. In \S 6, we draw our concluding remarks.

\section{Observation and Data Analysis}
We chose these four candidates (see Table 1 for details) for the 
availability of simultaneous radio and 
X-ray archival and published data. All of the four candidates have similar 
masses. Thus, the light crossing delay for all the sources are nearly same iff 
the disk size remained similar. Soft lags are mostly seen for candidates where 
inclination angle is high ($60^{\circ}<\theta<80^{\circ}$). The chosen candidates 
satisfy all of the above conditions.

\subsection{Radio Data}
We have used available archival Karl G. Jansky Very Large Array (VLA)\footnote{https://science.nrao.edu/facilities/vla/archive/index}{} data 
for our study. VLA has twenty seven fully steerable antennas of 25 diameter each, arranged in a `Y' shaped 
array. Antenna configurations depend on scientific goals. The most 
expanded configuration is known as the `A' configuration and the most compact one is called the 
`D' configuration. `B' and `C' configurations are intermediate. Occasionally antennas are 
placed in a hybrid configurations, like `AB', `BC' or `CD' when some of the antennas 
are in one configuration and some of them are in another. The maximum size of the baselines 
$B_\mathrm{max}$ in A, B, C and D configurations are 36.4, 11.1, 3.4 and 1.03 km respectively. 
For XTE 1859+226, the antenna were in AB configuration. The antenna were in A and D configuration 
for seven and twenty one days respectively for H1743--322. Here, we are only concerned about 
the integrated flux density of the object. So VLA configuration has no effect on the result.

Analysis and imaging of the data was carried out with Astronomical Image 
Processing System {\tt AIPS}\footnote{http://www.aips.nrao.edu}{}. 
The archival VLA data was read in AIPS using the task {\tt FILLM}. After 
checking the data, we have flagged out the bad data using the task {\tt SPFLG}.
Then the task {\tt SETJY} is used for providing the necessary source information.
We have determined the calibration of the data using the task {\tt CALIB}.
We use the task {\tt GETJY} to determine source flux densities.  
We used the flux density scale of \cite{Pe13}. After that, the task 
{\tt CLCAL} was used to apply the calibration solutions to the data.
Ten seconds of integration time was used for solving amplitude and phase calibration.
For the imaging and cleaning, we have used the task {\tt IMAGR} and 
make the final image of target source. The flux density is calculated using 
the task {\tt JMFIT} using the results of fitting a Gaussian, along with a background level.
More details about the VLA data reduction can be found in the VLA 
cookbook\footnote{http://www.aips.nrao.edu/cook.html}{}.
 
For XTE 1550-564, we have used the published radio flux at 843 MHz from MOST 
radio telescope by \cite{Wu02} and 8.6 GHz from ATCA by \cite{Ha01}. We  have 
also used the published radio flux at  8.3 GHz and 15.2 GHz for the source 
GRS 1915+105 from Green Bank Interferometer Monitoring Observations and Ryle 
radio telescope respectively by \cite{Mu01}.

\subsection{X-Ray Data}
We have produced time lag spectra for each observation using RXTE PCA archival data. 
The cross spectrum is calculated using $CF(j)=f_{1}^*(j) \times f_{2}(j)$, where $f_1$ 
and $f_2$ are the complex Fourier coefficients for the two energy bands at a frequency 
$\nu_j$. Here $f_1^*(j)$ is the complex conjugate of $f_1(j)$ \citep{van87}. The phase 
lag between two band signals at a Fourier frequency $\nu_j$ is given by $\phi_j = arg[CF(j)]$ 
and the corresponding time lag is $\phi_j/{2 \pi \nu_j}$ \citep{Ut14}. We calculated an average 
cross vector $CF$ by averaging complex values over multiple adjacent 16s cross 
spectra and then finding the final value of time lag versus frequency. We calculate 
time lags at the QPO centroid frequency ($\nu_c$) averaging over the interval $\nu_c \pm FWHM$ 
for 5-13 keV energy band against 2-5 keV energy band. For all cases of lag spectra, 
positive lag values mean that hard photons are lagging behind soft photons. We found that 
dead time effects are negligible and have ignored all the lag data where the error bar is more 
that $20\%$ of the time lag value. Current article contains only the lag properties of 
Low Frequency QPOs (type-A, type-B and type-C, see \cite{Mo15} for further details) 
within the frequency range of $0.01-15.0$ Hz.

\section{Results}

\begin{figure*}
\vskip 0.0cm
{\includegraphics[height=1.6\columnwidth,width=2.3\columnwidth]{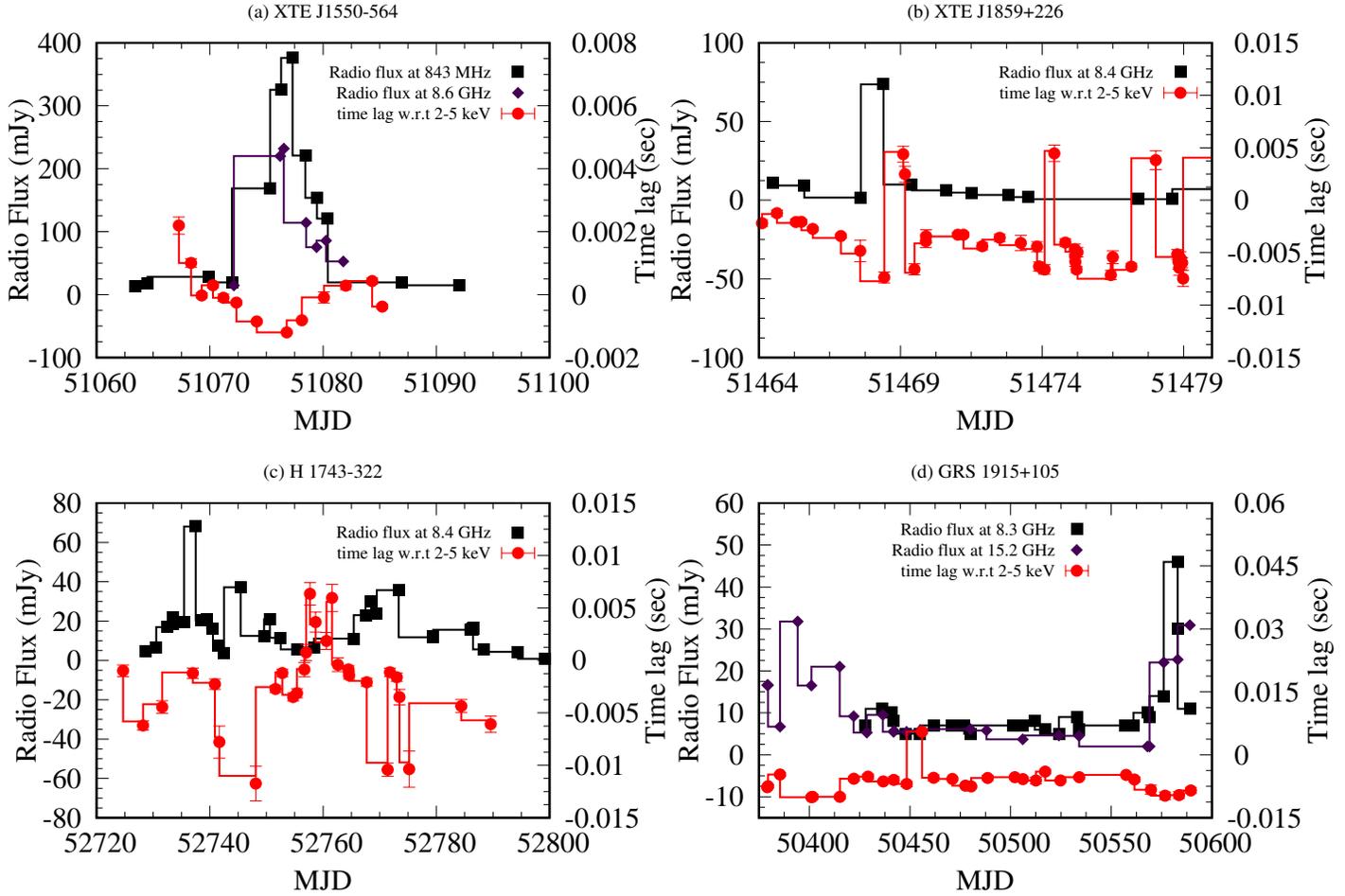}}\hskip 0.2cm
\caption{(a) Radio flux and X-ray soft lag shows perfect correlation for XTE J155-564. Observed radio maxima (in 8.6 GHz)
coincides with soft lag maxima. (b) Radio flux and X-ray soft lag correlation for XTE J1859+226 is shown. Hard lags
are only found during radio quite states. (c) Radio flux and X-ray soft lag correlation for H 1743-322 is presented. 15 days
are subtracted from the original X-ray data as radio leads X-ray for H 1743-322. (d) Radio flux and X-ray soft lag correlation
is shown for GRS 1915+105. {\bf Error bars, in some cases are smaller than the point size.}}
\label{}
\end{figure*}

\subsection{XTE J1550-564}
XTE J1550-564 was discovered by RXTE ASM \citep{Sm98}. The ASM count of the outburst 
achieved its peak on MJD 51076. The radio (\cite{Ca98}) and optical (\cite{Or98}) 
counterparts of this Galactic transient were detected shortly after that. 
Radio flux at 843 MHz \citep{Wu02} peaked roughly ($\sim 58$~hours) two days 
later than the X-ray. However, the maximum flux at higher frequencies \citep{Ha01} 
was seen a day before (see Fig. 2a). 

During the rising phase (51065-51076), QPO frequency is increased
monotonically (see, DC16) with a sudden `hiccup'
on MJD 51072. Declining phase started after MJD 51076 where the QPO frequency started to
decrease. Interestingly, we find that a soft lag maximum occurred on the same date 
(see Fig. 2a) as the radio peak flux in 8.6 GHz band. The days when the detected radio flux was low, 
the lag remained positive. With increasing radio flux, the magnitude of soft-lags started to increase.

\subsection{XTE J1859+226} 
\cite{Wo99} discovered XTE J1859+226 using on board ASM of RXTE mission. 
Radio \citep{Po99} and optical counterparts were discovered \citep{Ga99}.  
\cite{Mo15} classified the QPO types of this particular candidate along 
with several others according to their inclination. \cite{Br02} suggested 
disc-jet connection from the radio-X-Ray correlations for this particular outburst.

Within our region of interest, the source exhibited one giant radio flare and few short radio bursts. Out of all
the X-ray data that we have analyzed, positive lags are found in only eight days. During the 
peak of radio flux, we find a local soft lag minimum (see Fig. 2b). However, the lag sign switched 
soon after and simultaneous radio flux also decayed. Radio data during the interval
MJD $51468.5~\mathrm{to}~51469$ was unavailable (see Fig. 2b) which could have provided reason
for switching of lag sign. The regions where hard lags were detected (see Fig. 2b), observational 
data in nearby days showed that the object was in a rather radio quiet state.

\subsection{H 1743-322}   
Ariel V \citep{Ka77} and HEAO 1 \citep{Do77} satellites discovered H 1743-322 during
its 1977 outburst. After remaining in quiescent state for $26$ years, the source underwent a 
long outburst (discovered by \cite{Re03}) in 2003. The source was detected in wide range 
of electromagnetic spectrum (see \cite{Co05} for radio counterparts; \cite{St03} for 
optical counterparts). \cite{Mc09} performed extensive multi-wavelength studies of the
2003 outburst and discussed similarities with the 1998 outburst of XTE J1550-564.
However, the source showed anomalous behavior during its 2003 outburst 
(see \cite{Ch19} for details) and requires attention while understanding the 
accretion dynamics. Since then, the source exhibited series of outbursts around one or more per year.

VLA radio data (see, details in Table 1) during that outburst are analyzed.
During the 2003 outburst, the first radio peak occurred on MJD 52737. It is seen that 
the ASM count and radio profiles generally correlate if the X-ray data
is shifted to superimpose on radio emission date earlier by 15 days. 
We calculated Discrete Cross Correlation Function (DCCF) formulated by \cite{ek88}, 
which confirms a 15 days lag between radio and X-rays. \cite{Te19} found a delay of 
$12.0^{+3.7}_{-4.2}$ minutes between 24 GHz and 5 GHz for V404 Cygni during its 2015 outburst.
Given the energy range, one can find delays in the order of days between radio band and X-ray. 
A delay between radio and X-ray of 15 days is highly unlikely to be seen for other 
outbursting cases. But, it should not be neglected in the case of H 1743-322 during its 
2003 outburst.

\begin{figure} [h!]
\vskip 0.9cm
{\includegraphics[height=0.6\columnwidth,width=0.9\columnwidth]{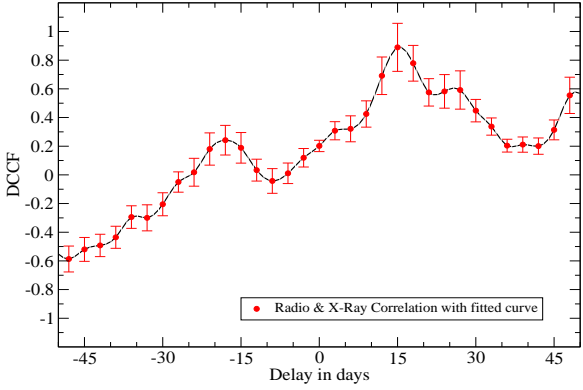}}\hskip 0.2cm
\caption{Discrete Cross Correlation Function (DCCF) is plotted for the radio and X-ray light curves
presented in Fig. 1c. DCCF maximizes on 15 days suggesting an X-ray delay.}
\label{}
\end{figure}

So, we performed the shift on the X-ray lag data following the delay observed in Fig. 1c. 
During this time, the source exhibited one giant ($\sim 70$ mJy) flare
and five smaller radio bursts. During the first radio peak, we see soft-lags as we have observed for
the previous two candidates. However, the magnitude of the lag is relatively small. 
Trailing that, rest of the radio maxima correlate (blue lines in Fig. 2c) 
with the observed soft-lag maxima. We also see the reverse correlation (between MJD 52750 to 52760)
where the hard lags are found in the absence of radio jets.

The reason for such kind of X-ray delay is poorly known and 
can be conjectured that the X-ray photons were trapped due to a sudden overflow 
of matter. However, further investigations over radio, optical, and X-ray bands are 
needed to specifically address the reason of radio lead for this atypical outburst.

\subsection{GRS 1915+105} 
\cite{Ca92} discovered GRS 1915+105 using WATCH on board GRANAT satellite. 
Apparent superluminal jets were detected (\cite{Mi94}). The candidate showed
complex X-ray flux variabilities both in long and short duration \citep{Gr96}.
For LFQPOs, \cite{Re00} reported both hard and soft lags. \cite{Mu01} 
performed multi-wavelength studies of GRS 1915+105 during
1996-1997 where no apparent correlation between X-ray and radio 
were reported.

During this period, around two radio flares are seen. Activity in radio is reduced substantially in the interval
of MJD 50450 to MJD 50550 (see Fig. 2d). We excluded the region between MJD 50250 to 50350 and beyond MJD 50600
due to the sparse sampling of the data. We can see a local radio maxima near about MJD 50400 where the local 
soft lag minima is also present. Another instance of the correlation can be found 
near about MJD 50575. We see the hard lags spiked during this interval while the source is in rather 
radio quite state and after that the lag sign switched again. \cite{Re00} pointed out the magnitude asymmetry of 
the hard and soft lags of GRS 1915+105.

\section{Correlation: Radio Flux vs Time Lag}
\begin{figure} [h!]
\vskip 0.5cm
{\includegraphics[height=0.7\columnwidth,width=1.0\columnwidth]{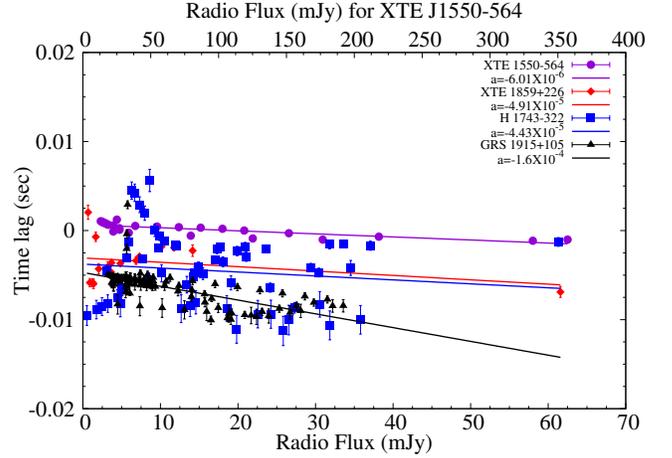}}\hskip 0.2cm
\caption{Correlation between radio flux and time lag is plotted for four objects. We opted for 
near simultaneous approach for the radio and time lag data. Each object was fitted with different 
slopes as properties of each objects may be vary. The gross slope was also found to be negative.}
\label{}
\end{figure}

The correlation between radio flux and time is plotted in Fig. 4. We use 
the least square fitting method to obtain the slope and intercept. 
We separately fitted each curves corresponding to a particular source. All of the 
fitted curves show negative slopes which implicates the increase of soft lag with 
the increase in radio flux. Best fitted curve was obtained for XTE J1550-564 with 
the least amount slope. The reason behind it might be relatively lesser magnitude 
of soft lag which was detected during its 1998 outburst. Steepest slope was observed
for GRS 1915+105. The reported lags (\cite{Mu01}, \cite{Du18}) for this candidate are 
much larger compared to others.

\subsection{Test of Significance}
We consider the null hypothesis 
$$
H_{0}=\mathrm{radio~jets~are~not~concurrent~with~soft~lags}
$$
and the 
alternative hypothesis 
$$
H_{a}=\mathrm{radio~jets~are~concurrent~with~soft~lags}.
$$
Our claim would strengthen if the null hypothesis is rejected.
Level of confidence ($C$) is set to be $0.95$~or~$95\%$. Accordingly, the test of 
significance $\alpha=1.0-C=0.05$. To establish the null hypothesis, we want to 
calculate the probability {\it (p-value)} of finding hard lags when radio flux 
$f_{\nu}\geq \bar{f_{\nu}^{h}}$ mJy which is essentially an ``one sided right tail test". So, 
$$
|Z| = \bigg|\frac{\bar{f_{\nu}^{h}} - \bar{f_{\nu}}}{(\sigma/\sqrt N)}\bigg|,
$$ 
where $\bar{f_{\nu}^{h}}$ is the average of radio flux on the days where
hard lags are seen, $\bar{f_{\nu}}$ is the global average of radio flux
of a given candidate, $\sigma$ is the standard deviation 
and $N$ is the sample size. Thus, the probability of getting hard lags for $f_{\nu}\geq \bar{f_{\nu}^{h}}$
can be obtained from the standard chart using
$$
P(Z:f_{\nu}\geq \bar{f_{\nu}^{h}}) = 1 - P(Z:f_{\nu}\le\bar{f_{\nu}^{h}}).
$$

\begin{table}[h!]
  \begin{center}
    \caption{Correlation Analysis.}
    \label{tab:table1}
    \begin{tabular}{c|c|c|c} 
      \textbf{Candidate Name} & \textbf{Pearson coefficient} & \textbf{Z score} & \textbf{$P(Z:f_{\nu}\geq \bar{f_{\nu}^{h}})$}\\
      \hline
	    XTE J1550-564 & -0.754 & 2.26 & 0.012\\
	    XTE J1859+226 & -0.673 & 2.14 & 0.016\\
	    H 1743-322    & -0.495 & 6.05 & $\le 0.0003$\\
	    GRS 1915+105  & -0.644 & 6.71 & $\le 0.0003$\\
      \hline
    \end{tabular}
  \end{center}
\end{table}

We provide a chart with $Z$ scores and corresponding $P(Z)$ value for 
each candidates.

\newpage

From the chart it is evident that $P(Z)\ll\alpha$. So, we can reject the null 
hypothesis.

\section{Time Lag simulation}
\begin{figure} [h!]
\vskip 0.8cm
{\includegraphics[height=0.6\columnwidth,width=1.0\columnwidth]{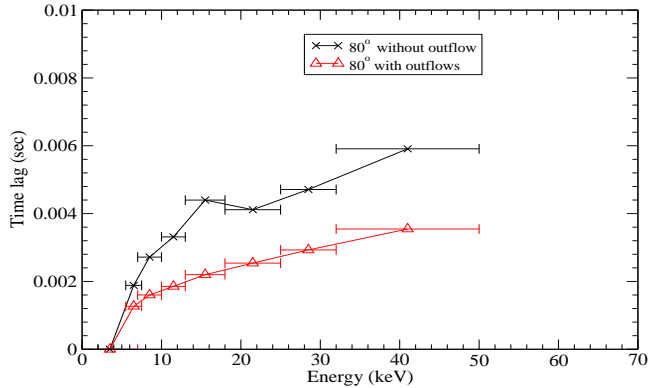}}\hskip 0.2cm
\caption{Simulated energy dependent time lags with and without the effect of outflows are presented for 
high inclination sources. The reference or soft band is 2.0 to 5.0 keV for simulated lag calculations.
In presence of outflows, time lag magnitude decreases. Currently, simulation
boundary extends up to 100 $r_g$ which is also the outer boundary of Keplerian disk. Outer boundary of 
Compton cloud is $~45~r_g$.}
\label{}
\end{figure}

To mimic the features of an outburst, energy dependent time lags were simulated (CCG17b) considering 
relativistic thick disks as Compton cloud with decreasing outer boundary. The Keplerian disk acts as 
a soft photon source whose exterior was located at $100~r_g$. The outflow part was neglected (CCG17b)
at first to simulate the lag spectra using anisotropic electron energy distribution in the Compton 
cloud. However, hydrodynamic simulation of sub-Keplerian flow allowed \cite{Ch18} to show time dependent images 
of disk-jet in presence of self-consistently produced outflows from Compton cloud. For similar outer 
boundary of the Compton cloud, the lag magnitude is found to be reduced (Fig. 5) for high 
inclination angle sources where outflows are considered.

\section{Discussions: Possible Role of Outflows}
One can infer a possible correlation between the soft lag and radio flux profiles from Fig. 2.
To understand this behavior, we use two component advection flow solution 
which is based on transonic flow solutions around black holes (see CT95 for details).
The model contains a high viscosity Keplerian disc on the equatorial plane which is surrounded by a
low angular momentum advective component which produces a centrifugal barrier close to the black hole 
horizon. Numerical simulations of this advective component shows presence of strong
winds \cite{Mo94} from the Compton cloud or CENtrifugal pressure supported BOundary 
Layer also known as CENBOL. Extending this to full general relativistic flow, recently
\cite{Ki19} showed that most of the outflow has sub-escape velocity and 
returns back to the equatorial region while only a fraction can achieve escape 
velocity and leave the system along the axis from Compton cloud. We conjecture 
that this return flow down-scatters hard photons and creates
soft photons with large time lag. In principle, this should affect the 
time lag spectra of black holes. Since return flows are always associated with 
the jet activity, we find soft-lags to be higher when activity of the jets is 
also high. Figure 6 presents the cartoon diagram of the system. The effects 
mentioned above are only discernible for high inclination angle systems as gravitational 
bending enhances the number of hard photons (see CCG17a) which will eventually 
downscatter with the Returing OutFlows (ROF) along the line of sight of the observer.

\begin{figure*}
\centering {
\hskip 1.5cm\includegraphics[height=0.70\columnwidth,width=1.5\columnwidth]{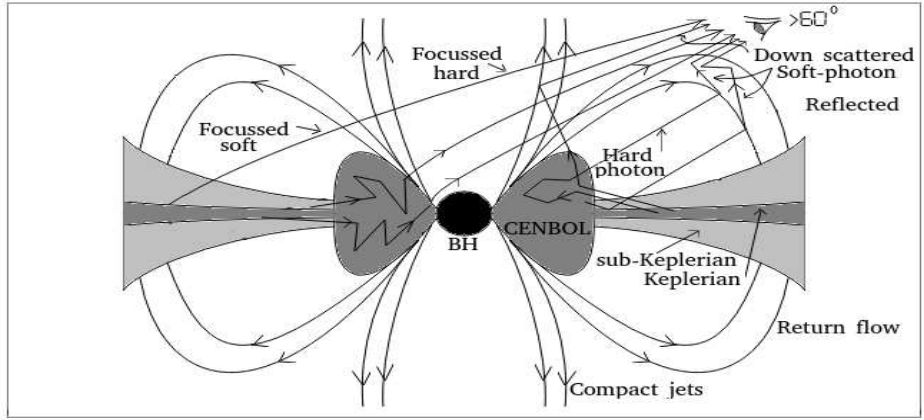}}
\caption{A cartoon diagram of the formation of jets and return flows from an advective disk
is presented. Counting from the left side, six types of emergent photons are considered here. The 
first three photons are coming from the other side of the accretion disk and have suffered gravitational
bending. Among them, the first one contributes in soft X-ray as it is not intercepted by the Compton cloud 
while the other two are inverse Comptonized. The fourth one is reflected and downscattered 
photon in the ``ROF" region while the rest of the two types are hard photons 
which are downscatted in ``ROF" and compact jet regions respectively.}
\label{Cartoon}

\end{figure*}

Also, there may be other physical processes like lense-thirring precision \citep{in09}, 
disk reflection \citep{Ka13} which are proposed to be the contributor of the soft lags.
Recently, \citep{Sc17} a correlation between radio luminosity and measured 
spin of the Quasars suggested that the highly spinning black holes are more prone 
to be radio loud. This directly connects if soft lags were generated due to the spin of 
the central objects and their radio intensities. However, within our current studies, 
XTE J1550-564 showed most promising correlation between soft lags and radio intensities 
and the reported \citep{St11} spin for that candidate is $0.49\pm0.24$ indicating it 
to be an intermediate spin black hole. 
Using the Lamp-Post model proposed by \cite{MM96} and \cite{MF04}, the effect of outflows 
on lag behavior can also be understood. Outflow maximizes during the intermediate state 
where X-ray flux higher than the low state. During outflows (assuming the magnetic field 
of the lamp post corona drives out matter isotropically), number of photons reflected
by the disk increases which could also enhance the soft lag magnitude (\cite{Ka13}, see \S 4).
This could also explain the possible correlation between outflows and soft lags. Recently, 
\cite{Re18} showed correlations between photon index and lag magnitude. Their 
simulation (\cite{RK19}) shows a correlation of the base of the jet and lag magnitude which 
can be interpreted as the change in spectral state. It would be interesting investigate such 
correlation changing the energy ($\varepsilon$), angular momentum ($\lambda$) and accretion rate ($\dot{m}$) 
(see for details \cite{Ch90}) in the hydrodynamic simulations of sub-Keplerian flow. But, the work is 
beyond the scope of this paper and will be reported elsewhere. 

\newpage

\section{Conclusions}
X-ray lags are important to understand the evolution of the geometry of an accretion disk and
dominant physical mechanism. We showed here that soft lags are correlated with the radio intensity 
for some low mass X-ray binaries. We believe that this is a generic property of such systems 
since whether jets are produced ahead of outbursts (as in H1743-322) or during the outbursts, return flows
will always downscatter hard photons as observed by high inclination systems. We conclude that:

\begin{enumerate}
\item []1. Soft-lags are concurrent with the radio flares. 
\item []2. Outflows evolve with the evolution of disk. Along with Comptonization, reflection and effect of
curved geometry (see DC16 and CCG17b), return outflows play a crucial role in the evolution of lag profiles.  
\item []3. Our result demonstrates a model independent physical phenomenon which connect disk
properties with jet activities.
\end{enumerate}

Radio and X-ray time lag studies of a few GBHs at high inclination angle revealed new features in disk-jet 
paradigm. Recent works of \cite{Mu01}, \cite{Ga08}, \cite{AM15}, \cite{Ve17}, \cite{Ma18} and \cite{Te19} have 
shown promising results on jet involvement with previously predicted disk activities. So, to understand more 
about the lag features and their origin, one must emphasize on the simultaneous broadband observations of 
black hole binaries.

\section{Acknowledgements}
AC acknowledges Post-doctoral fellowship of S. N. Bose National Centre for Basic Sciences and 
Indian Centre for Space Physics where a part of this work was carried out. BGD acknowledges IUCAA
associateship. Authors thank Tomaso Belloni for providing the timing analysis software GHATS and 
anonymous Reviewer for suggestions and comments which enhanced the quality of the manuscript.

{}


\end{document}